\documentclass[prl,twocolumn,showpacs,]{revtex4-1}
\usepackage{amsmath}
\usepackage{amssymb}
\usepackage[dvips]{graphicx}
\usepackage{dcolumn}
\usepackage{color}
\graphicspath{{images/}}
\begin{document}

\title{Spin-orbit coupling and optical spin Hall effect in photonic graphene}
\author{A. V. Nalitov}
\author{G. Malpuech}
\author{H. Ter\c{c}as}
\author{D. D. Solnyshkov}

\affiliation{Institut Pascal, PHOTON-N2, Clermont Universit\'{e}, Blaise Pascal
University, CNRS, 24 avenue des Landais, 63177 Aubi\`{e}re Cedex, France.}

\begin{abstract}
We study the spin-orbit coupling induced by the splitting between TE and TM optical modes in a photonic honeycomb lattice. Using a tight-binding approach, we calculate analytically the band structure. Close to the Dirac point,we derive an effective Hamiltonian. We find that the local reduced symmetry ($\mathrm{D_{3h}}$) transforms the TE-TM effective magnetic field into an emergent field with a Dresselhaus symmetry. As a result, particles become massive, but no gap opens. The emergent field symmetry is revealed by the optical spin Hall effect.

\end{abstract}

\maketitle

Spin-orbit coupling in crystals allows to create and control spin currents without applying external magnetic fields. These phenomena have been described in the seventies \cite{Dyakonov} and are nowadays called the spin Hall effect (SHE) \cite{Hirsch1999, reviewSHE}. In 2005, the interplay between the spin-orbit coupling and the specific crystal symmetry of graphene\cite{Geim2007} has been proposed \cite{Kane2005} to be at the origin of a new type of spin Hall effect, the quantum spin Hall effect, in which the spin currents are supported by surface states and are topologically protected \cite{QSHE,Kane2010}. This result has a special importance, since it defines a new class of $Z_2$-topogical insulator \cite{Kane2005b}, not associated with the quantization of the total conductance, but associated with the quantization of the spin conductance. However, from an experimental point of view, the realization of any kind of SHE is difficult, because spin-orbit coupling does not only lead to the creation of spin current, but also to spin decoherence \cite{Dyakonov2}. In graphene, the situation is even worse, since the spin-orbit coupling is extremely weak. Deposition of adatoms has been proposed to increase the spin-orbit coupling \cite{Gmitra2013}, and  it allowed the recent observation of the SHE \cite{Balakrishnan2013}, but associated with a very short spin relaxation length, of the order of 1 $\mu$m.

On the other hand, artificial honeycomb lattices for atomic Bose Einstein Condensates (BEC) \cite{hon_atom} and photons \cite{Peleg2007,Kuhl2010,Polini2013,Kalesaki2014,Jacqmin2014} have been realized. These systems are gaining a lot of attention due to the large possible control over the system parameters, up to complete Hamiltonian engineering\cite{Hafezi,Umucalilar}. In BECs, the recent implementation of synthetic magnetic fields \cite{Lin1} and of non-Abelian, Rashba-Dresselhauss gauge fields \cite{Lin2} appears promising in the view of the achievement of topological insulator analogs.  Photonic systems, and specifically photonic honeycomb lattices appear even more promising. They are based on coupled wave guide arrays \cite{Rechtsman2013}, on photonic crystals with honeycomb symmetry \cite{Won2011}, and on etched planar cavities \cite{Jacqmin2014}. A photonic Floquet topological insulator has been recently reported \cite{Chong2013}, and some others based on the magnetic response of metamaterials predicted \cite{Khanikaev}. In photonic systems, spin-orbit coupling naturally appears from the energy splitting between the TE and TM optical modes and from structural anisotropies. Both effects can be described in terms of effective magnetic fields acting of the photon (pseudo)-spin \cite{Shelykh2010}. In planar cavity systems, the TE-TM effective field breaks the rotational symmetry, but preserves both time reversal and spatial inversion symmetries. It is characterized by a $k^2$ scaling and a double azimuthal dependence. This spin-orbit coupling is at the origin of the optical spin Hall effect (OSHE)\cite{Kavokin2005,Leyder2007} and of the acceleration of effective magnetic monopoles \cite{Hivet,Bramwell2012,Solnyshkov2013}. As recently shown \cite{Tercas2014}, the specific TE-TM symmetry can be locally transformed into a non-Abelian gauge field in a structure with a reduced spatial symmetry.

In this work, we calculate the band structure of photonic graphene in the presence of the intrinsic spin-orbit coupling induced by the TE-TM splitting. We derive an effective Hamiltonian which allows to extract an effective magnetic field acting on the photon pseudo-spin only. We find that the low symmetry ($\mathrm{D_{3h}}$) induced by the honeycomb lattice close to the Dirac points transforms the TE-TM field in a emergent field with a Dresselhaus symmetry. Particles become massive but no gap opens. The dispersion topology shows larges similarities with the one of bilayer graphene \cite{McCann2006} and of monolayer graphene with Rashba spin-orbit coupling \cite{Rakyta2010}, featuring trigonal warping \cite{Dresselhaus1974} and Lifshitz transition \cite{Lifshitz1960}. The symmetry of these states is revealed by the optical spin Hall effect (OSHE) which we describe by simulating resonant optical excitation of the $\Gamma$, K and K' points. The OSHE at the $\Gamma$ point shows four spin domains associated with the TE-TM symmetry. The OSHE at the K and K' shows two domains characteristic of the Dresshlauss symmetry. The spin domains at the K and K' points are inverted, which is a signature of the restored $\mathrm{D_{6h}}$ symmetry when the two valleys are included. 
  
In what follows, in order to be specific, we consider a honeycomb lattice based on a patterned planar microcavity similar to the one recently fabricated and studied \cite{Jacqmin2014}. This does not reduce the generality of our description, which can apply to other physical realizations of honeycomb lattices, in optical and non-optical systems. In \cite{Jacqmin2014}, quantum wells were embedded in the cavity which provided the strong coupling regime and the formation of cavity exciton-polaritons. Here, we will consider the linear regime, a parabolic in-plane dispersion, and no applied magnetic field. In such case, photons and exciton-polaritons behave in a similar way and our formalism applies to both types of particles.

\begin{figure} \label{fig1}
\includegraphics[scale=0.215]{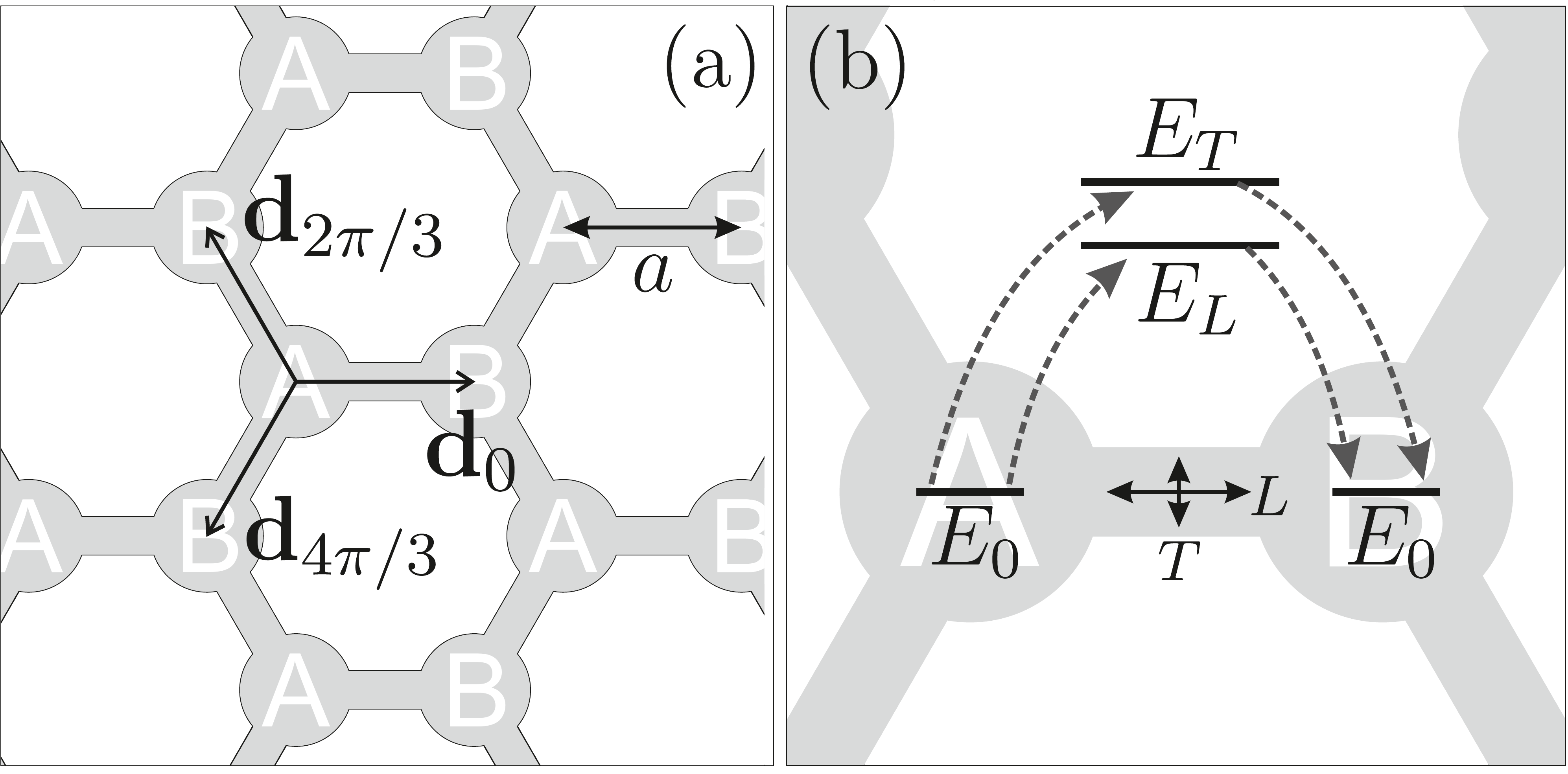} 
\caption{(color online) A schematic sketch of the tight-binding model. (a) Photon tunneling between microcavity pillars is described as photon propagation through "waveguide"-like links. (b) Polarization dependence of tunneling probability due to TE-TM energy splitting: $L$ state which is polarized longitudinally to the "waveguide" link is closer in energy to the degenerate pillar-pinned states than the transversely-polarized state $T$, resulting in higher L-photon tunneling probability through the link. }
\end{figure} 

%
%

%
\emph{Tight-binding model} 
First, we describe the spin-orbit coupling in photonic graphene structure (figure 1a) within the tight-binding approximation.
We take a basis of $\sigma\pm$ polarized photon states localized on each pillar of the lattice as a zeroth approximation for the tight-binding model and introduce the hopping of photons from a pillar to one of its nearest neighbors as a perturbation $\hat{V}$ on this basis.

To illustrate the polarization dependence of the hopping probability, let us consider two neighbouring pillars $A$ and $B$, shown in Figure (1b).
The photon hopping between them may be described as propagation through a "waveguide"-like link.
TE-TM energy splitting imposes a slight difference $\delta J$ in tunneling matrix elements for states linearly-polarized longitudinally ($L$) and transversely ($T$) to vector $\mathbf{d}_\varphi$ linking the pillars \cite{suppl}, as it was recently shown for the eigenstates in a photonic benzene molecule \cite{Vera}. In that framework, the matrix elements read:
\begin{equation}
\langle A, L \vert \hat{V} \vert B, L\rangle \equiv -J-\delta J/2, \quad
\langle A, T \vert \hat{V} \vert B, T\rangle \equiv -J+\delta J/2. \notag
\end{equation}
While a photon is in a link, TE-TM field does not rotate its eigenstate polarizations $L$ and $T$, implying no cross-polarization matrix elements:
\begin{equation}
\langle A, L \vert \hat{V} \vert B, T\rangle = \langle A, T \vert \hat{V} \vert B, L\rangle = 0. \notag
\end{equation}
In $\sigma\pm$ basis, the probability of spin flip during hopping is linear in $\delta J$ and its phase gain depends on the  angle $\varphi$ between the link and the horizontal axis:
\begin{equation}
\langle A, \pm \vert \hat{V} 
\vert B, \pm \rangle = -J, \quad
\langle A, + \vert \hat{V} 
\vert B,- \rangle =
- \delta J e^{-2\mathrm{i}\varphi}. \notag
\end{equation}
This phase factor reflects the fact that when a link is rotated by 90 degrees, $L$ and $T$ polarization basis is inverted: if $L$ was horizontal, it becomes vertical and vice versa.

A photon state may be described in the bispinor form $\Phi = \left( \Psi_A^+, \Psi_A^-, \Psi_B^+, \Psi_B^- \right)^{\mathrm{T}}$, with $\Psi_{A(B)}^\pm$ being the wave function on both sublattices in both spin components.
The effective Hamiltonian acting on a plane wave bispinor $\Phi_\mathbf{k}$ then has a block matrix form:
\begin{equation} \label{Hamiltonian}
\mathrm{H}_\mathbf{k} = \left( \begin{matrix}
\mathrm{0} & \mathrm{F}_{\mathbf{k}} \\
\mathrm{F}_{\mathbf{k}}^\dagger & \mathrm{0}
\end{matrix} \right), \quad
\mathrm{F}_{\mathbf{k}} = - \left( \begin{matrix}
f_{\mathbf{k}} J & f_{\mathbf{k}}^+ \delta J \\
f_{\mathbf{k}}^- \delta J & f_{\mathbf{k}} J
\end{matrix} \right),
\end{equation}
where complex coefficients $f_{\mathbf{k}}$,$f_{\mathbf{k}}^\pm$ are defined by:
\begin{equation}
f_{\mathbf{k}}=\sum_{j=1}^3 \exp(-\mathrm{i}\mathbf{k d}_{\varphi_j}),\quad
f_{\mathbf{k}}^\pm = \sum_{j=1}^3 \exp(-\mathrm{i}\left[\mathbf{k d}_{\varphi_j} \mp 2 \varphi_j \right]), \notag
\end{equation}
and $\varphi_j = 2 \pi (j-1) / 3$ is the angle between the horizontal axis and the direction to the $j$th nearest neighbor of a type-A pillar.
Its diagonalization results in a biquadratic equation on the photon dispersion, having two pairs of solutions $\pm E_\mathbf{k}^\pm$, given by:
\begin{align}\label{disp_sol}
&2  (E_\mathbf{k}^\pm)^2 =
2 \vert f_\mathbf{k} \vert^2 J^2 +
\left( \vert f_\mathbf{k}^+ \vert^2 + \vert f_\mathbf{k}^-\vert^2 \right)
\delta J^2 \pm \\ 
&\pm \sqrt{(\vert f_\mathbf{k}^+ \vert^2 - \vert f_\mathbf{k}^- \vert^2)^2 \delta J^4 + 4 \vert f_\mathbf{k} f_\mathbf{k}^{+*}+f_\mathbf{k}^* f_\mathbf{k}^- \vert ^2 J^2 \delta J^2}. \notag
\end{align}

The dispersion is plotted along the principal direction in Figure (2a), and the trigonal warping effect which is a characteristic of bilayer graphene \cite{McCann2006} and of monolayer graphene with Rashba spin-orbit coupling \cite{Rakyta2010} is shown on the Figure (2b) in the vicinity of the K point. When $\delta J = J /2$, trigonal warping disappears. The crossing points originating from different Dirac points meet and annihilate. The dispersion topology changes -- a phenomenon associated with the so-called Lifshitz transition in Fermionic systems \cite{Lifshitz1960}.%

If $\delta J \ll J$, the distance $\delta K$ between a K point and the additional pockets is approximately given by $(\delta J / J)^2 a^{-1}$.

\begin{figure} \label{fig2}
\includegraphics[scale=0.4]{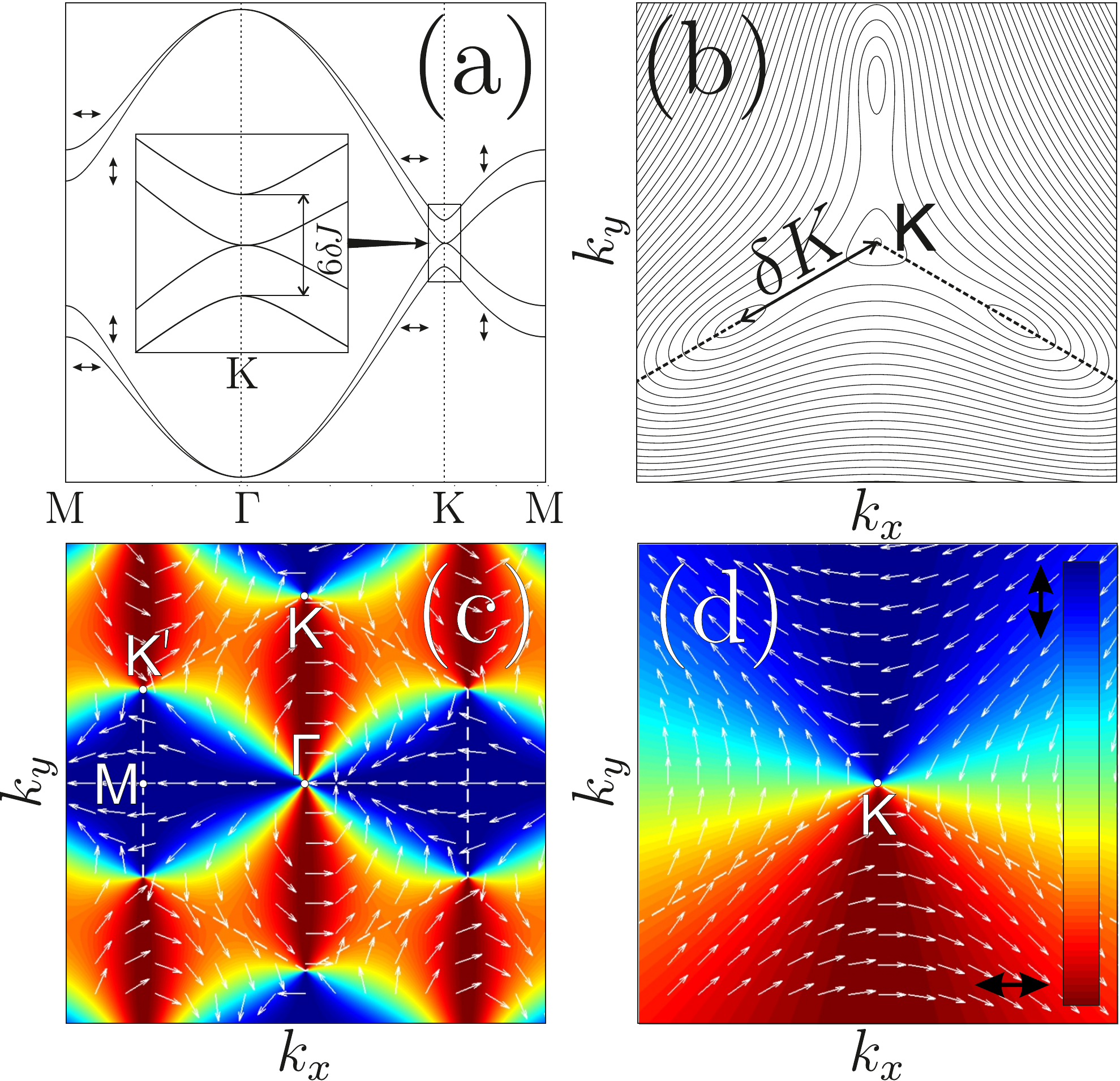} 
\caption{(color online) The eigenstates of the effective Hamiltonian due to the honeycomb potential and the TE-TM splitting. (a) Photon dispersion branches along principal directions. The inset demonstrates a zoomed K valley -- the dispersion is gapless. (b) Isoenergetic lines around K point, illustrating the trigonal warping: Dirac cones split in four pockets. (c) and (d) Linear polarization map and the pseudospin texture of the lowest energy state.}
\end{figure}

The effective Hamiltonian may be expressed in terms of pseudospin operators $\boldsymbol{\sigma}$ and $\mathbf{s}$, having the same matrix form of Pauli matrices vector and corresponding to sublattice (A/B) and polarization (H/V) degrees of freedom.
It may further be separated into polarization-independent part $H^{(0)}_\mathbf{k}$, coupling $\boldsymbol{\sigma}$ with momentum and giving a standard graphene dispersion with two Dirac valleys K and K$^\prime$, and a spin-orbit term $H^{\mathrm{SO}}_\mathbf{k}$, coupling $\mathbf{s}$ with $\boldsymbol{\sigma}$ and momentum:
\begin{align}
H_\mathbf{k}^{\mathrm{(0)}} =& - J \sigma_+ f_\mathbf{k} +h.c., \label{H_0}\\
H_\mathbf{k}^{\mathrm{SO}} =& - \delta J \sigma_+ \otimes \left( f_\mathbf{k}^+ s_+ + f_\mathbf{k}^- s_- \right) + h.c.. \label{H_SO}
\end{align}
where $\sigma_\pm = (\sigma_x \pm \mathrm{i} \sigma_y)/2$, $s_\pm = (s_x \pm \mathrm{i} s_y)/2$, and the $\otimes$ symbol denotes Kronecker product.
Expanding expressions (\ref{H_0},\ref{H_SO}) and keeping the main order in $\mathbf{q=k-K}$, we further isolate the momentum-independent part $H^\mathrm{SO}_{\mathbf{K}}$ coupling $\mathbf{s}$ with $\boldsymbol{\sigma}$ and rewrite both terms in the low-energy approximation:
\begin{align}
H_\mathbf{q}^{\mathrm{(0)}} =& \hbar v_F \left( \tau_z q_x \sigma_x + q_y \sigma_y \right), \label{H_01} \\
H^{\mathrm{SO}} =& \Delta \left( \tau_z \sigma_y s_y - \sigma_x s_x \right), \label{H_SOK}
\end{align}
where $v_F = 3 J a / (2 \hbar)$, $\Delta = 3 \delta J / 2$ and $\tau_z$ equals $+1$ and $-1$ for K and K$^\prime$ valleys respectively. 
Here we use the same basis as the the one of Kane and Mele \cite{Kane2005} in order to allow for a direct comparison with their Hamiltonian. This basis is different from the original basis of Wallace \cite{Wallace} which is used in the eq.\ref{Hamiltonian}. The passage from Wallace to Kane is obtained by writing $q_x\rightarrow q_y$,$q_y\rightarrow -q_x$).

If one restricts state space by locally fixing the sublattice $\boldsymbol{\sigma}(\mathbf{k})$ pseudospin and valley $\tau_z$, the spin-orbit term may be treated as an interaction with an emergent field.
As an example, if one considers eigenstates of the main term (\ref{H_01}) in one Dirac valley with a fixed energy sign $c=\pm 1$, spin-orbit term (\ref{H_SOK}) transforms to a symmetry-allowed  Dresselhaus-like emergent field:
\begin{align}
H_{c}^{\mathrm{SO}} =& -\Delta c \left( q_x s_x + q_y s_y\right) /q. \label{H_SOKt}
\end{align}

This term, having a well-defined physical origin, is similar in spirit with the Rashba term introduced by Kane and Mele \cite{Kane2005,Kane2005b}.  
The effective field described by the spin-orbit term (\ref{H_SOKt}) splits the degenerate massless photon branches by $3 \delta J$, and their linear polarization only depends on the direction of $\mathbf{q}$ and not on its absolute value.
However, if $q < \Delta / \hbar v_F = (\delta J / J) a^{-1}$, the spin orbit term cannot be considered as a perturbation of the main term (\ref{H_01}), the interplay between the two terms gives an effective photon mass $m^*=(2 c \hbar^2 \delta J)/(3 a^2 J^2)$ in this region of reciprocal space.
The pseudospin pattern (defining the linear polarization of light) of the lowest energy eigenstate reflects the effective field acting on the particles, because the pseudospin aligns with this field. The the whole reciprocal space is shown in Figure (2c). The figure (2d) shows a zoom on the K-point where the emergent Dressehlaus like field is clearly identified. Figure (2c) also clearly shows that the effective fields have an opposite sign close to the K and K' points respectively. 

From this analytical calculation of the dispersion, we can conclude that the particular type of spin-orbit coupling we consider does not open a gap in the K point of the Brillouin zone, but leads to the appearance of massive particles. This, among other consequences, should induce a strong modification of the Klein tunneling effect. As shown in \cite{Liu2012}, where Klein tunneling in the presence of a Rashba term was  considered, the tunneling is suppressed for energies close to the K point, where the dispersion is not linear anymore, but is recovered for higher energies.

The best evidence of the presence of a spin-orbit coupling inducing an effective magnetic field of a specific symmetry is the optical spin-Hall effect: rotation of the particle spin around the effective wavevector-dependent field during their propagation. The resonant excitation around the $\Gamma$ point with linearly polarised light should lead to a radial expansion of the wave-packet accompanied by a precession of the photon pseudo-spin. The double azimuthal dependence of the effective field orientation should lead, as in the planar case, to the formation of four spin domains \cite{Kavokin2005,NOSHE}. Close to the K and K' points, the Dressehlaus effective field orientation follows the azimuthal angle and two spin domains only should form \cite{Vishnevsky2013,Tercas2014}.  

\emph{Numerical simulation} In the following, in order to check the validity of the tight-binding approximation, and the observability of the OSHE, in realistic structures and experiments, namely including the broadening induced by the finite life time, we study numerically the propagation of polarised light in the photonic graphene structure. We consider a structure etched out of a planar microcavity, where the graphene atoms are represented by overlapping pillars (fig 3a). The equation of motion for the photonic spinor wavefunction reads :

\begin{eqnarray}
& i\hbar \frac{{\partial \psi _ \pm  }}
{{\partial t}}  =  - \frac{{\hbar ^2 }}
{{2m}}\Delta \psi _ \pm   + U\psi _ \pm   - \frac{{i\hbar }}
{{2\tau }}\psi _ \pm   + \\
& + \beta {\left( {\frac{\partial }{{\partial x}} \mp i\frac{\partial }{{\partial y}}} \right)^2}{\psi _ \mp } 
+P_0 e^{ { - \frac{{\left( {t - t_0 } \right)^2 }}
{{\tau _0^2 }}}}e^{ { - \frac{{\left( {{\mathbf{r}} - {\mathbf{r}}_0 } \right)^2 }}
{{\sigma ^2 }}}}e^{ {i\left( {{\mathbf{kr}} - \omega t} \right)} } \notag
\end{eqnarray}

where $\psi(r)={\psi_+(r), \psi_-(r)}$ are the two circular components of the photon wave function, $m$ is the cavity photon mass, $\tau$ the lifetime. This equation is similar with the one describing the photon motion in a planar cavity in the presence of TE-TM splitting \cite{Shelykh2010}, described by the parameter $\beta ={%
\hbar ^{2}}\left( {m_{l}^{-1}-m_{t}^{-1}}\right) /4m$ where $%
m_{l,t}$ are the effective masses of TM and TE polarized particles
respectively and $m=2\left( {{m_{t}}-{m_{l}}}\right) /{m_{t}}{%
m_{l}}$. We have taken $m_t=5\times10^{-5}m_0$, $m_l=0.95m_t$, where $m_0$ is the free electron mass. The only difference lies in the introduction of the honeycomb lattice potential $U(r)$ shown on the figure 3a (24x24 elementary cells).
$P_{0}$ is the amplitude of the pulsed pumping (identical for both components, corresponding to horizontal polarization), the pulse duration is $\tau_0=1$ ps, the size of the spot $\sigma=15$ $\mu$m. Pumping is localized in real space and in reciprocal space close to the selected point ($\Gamma$, K or K'). The lifetime was taken $\tau=25$ ps.

\begin{figure} \label{fig3}
\includegraphics[scale=0.4]{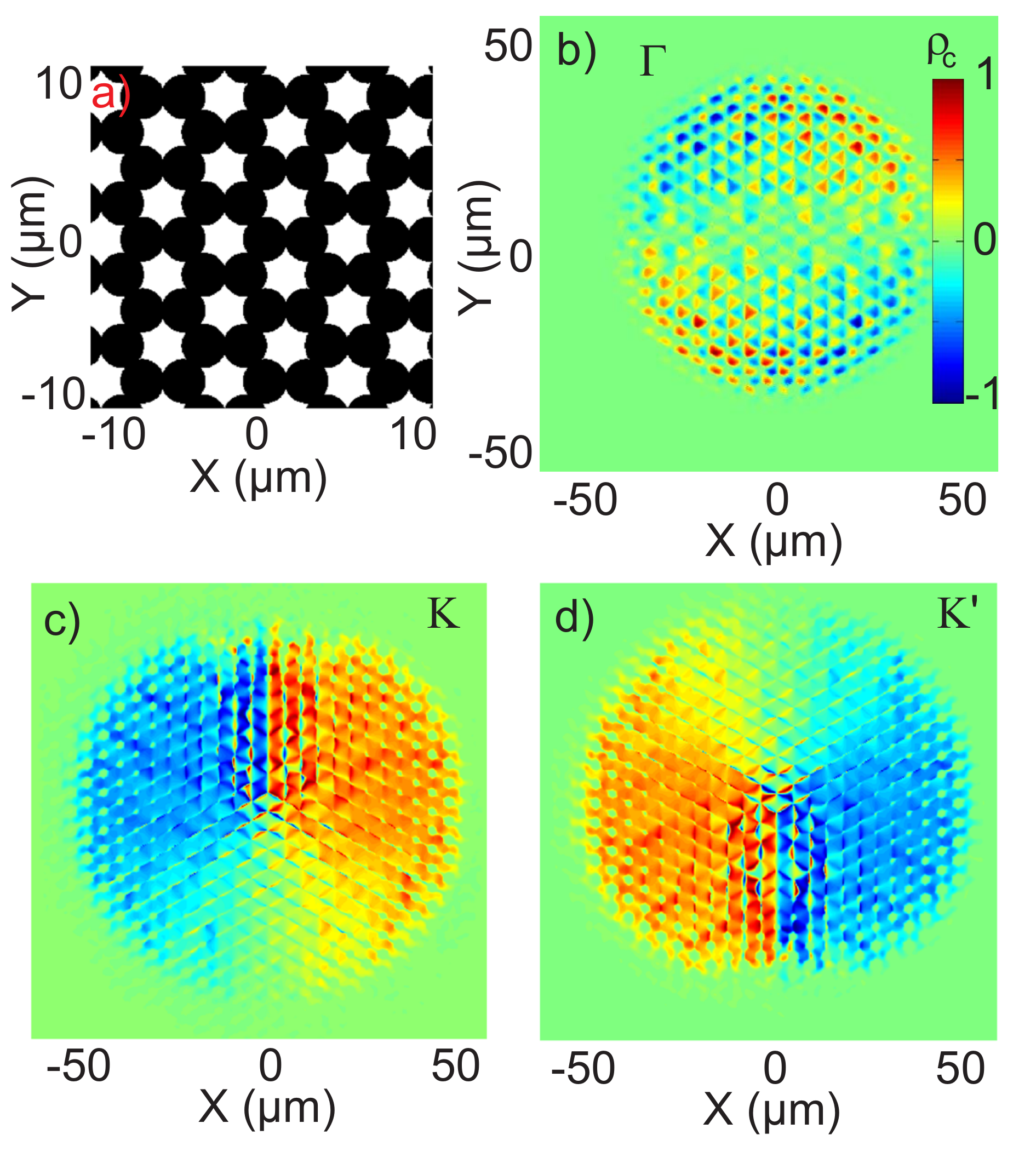} 
\caption{(color online). Optical spin Hall effect in photonic graphene. Circular polarization degree as a function of coordinates: a) the potential used in the simulations; b) excitation at $\Gamma$ point (TE-TM field); c) excitation at K point (Dresselhaus effective field); d) excitation at K' point (field inverted with respect to K').}
\end{figure} 

We have performed numerical simulation of optical spin Hall effect in photonic graphene using a high-resolution (512x512) representation of a potential, similar to the one already studied in experiments \cite{Jacqmin2014}. The nVidia CUDA graphical processor was used to carry out the integration of the 2D spinor Schroedinger equation.

Figure (3-b,c,d) shows the snapshots taken at $t=30 ps$ of the circular polarization degree as a function of coordinates. Panel b) shows the polarization degree for the excitation in the $\Gamma$ point, where the field has the typical TE-TM texture, evidenced by the 4 polarization domains \cite{Kavokin2005,Leyder2007,NOSHE}. Panel c,d) demonstrate the optical spin Hall effect for the $K$ and $K'$ points respectively, where the field has the texture of the Dresselhaus spin-orbit coupling. This is evidenced by 2 polarization domains in real space \cite{Vishnevsky2013,Tercas2014} being inverted between the $K$ and $K'$ points which reflects the fact that the fields around $K$ and $K'$ are respectively opposite. 

The texture of the optical spin-Hall effect is a clear demonstration of the different nature of the effective magnetic field due to the spin-orbit coupling in the two Dirac points (K and K') of the Brillouin zone. From this numerical experiment, we clearly see the advantage of photonic systems, which allow to excite and analyze any point of the dispersion, much easier than in solid state systems.

Other very interesting consequences of our work rely on the possibilities offered by the manipulation of the lattice geometry in photonic systems and by the mixed exciton-photon nature of exciton-polaritons. The system geometry is the tool which has been used to create a photonic topological insulator \cite{Rechtsman2013}. Combined with spin-orbit coupling, it opens very broad perspectives. The mixed nature of exciton-polaritons provides a magnetic response of the system at optical frequencies, which is of interest to realize a photonic topological insulator \cite{Haldane2008,Soljacic09}. It also induces a very strong non-linear optical response. Non-linear spin Hall effect associated with the  transmutation of topological defects and focusing of spin currents have been already described in planar structures \cite{NOSHE}. The behaviour of soliton states in photonic topological insulators was  recently considered \cite{Segev}. More generally, the interactions allow an exciton-polariton gas to behave as a quantum fluid \cite{Carusotto2013} with spin-anisotropic interactions \cite{Shelykh2010}. Polaritonic graphene \cite{Jacqmin2014}  therefore opens very large possibilities for the studies of interacting spinor quantum fluids, in the presence of different types of real and effective magnetic fields which suggest accessibility to different types of quantum phases.

To conclude, we have studied the spin-orbit coupling induced by the TE-TM splitting in a microcavity etched in the shape of a graphene lattice. Within the tight-binding approximation we found the eigenstates of the system, derived an effective Hamiltonian and found the effective fields acting on the photon spin. The symmetry of the field is lowered close to the Dirac points where it takes the form of a Dressehlauss field. The experimental observability of the optical Spin Hall effect induced by this spin-orbit coupling is verified by numerical simulations.

We acknowledge discussions with M. Glazov, A. Amo, I. Carusotto, and J. Bloch. This work has been supported by the ITN INDEX (289968), ANR Labex GANEX (Grant No. ANR-11-LABX-0014), ANR Quandyde (ANR-11-BS10-001) and IRSES POLAPHEN (246912).

\bibliography{reference}

\begin{thebibliography}{46}%
\makeatletter
\providecommand \@ifxundefined [1]{%
 \@ifx{#1\undefined}
}%
\providecommand \@ifnum [1]{%
 \ifnum #1\expandafter \@firstoftwo
 \else \expandafter \@secondoftwo
 \fi
}%
\providecommand \@ifx [1]{%
 \ifx #1\expandafter \@firstoftwo
 \else \expandafter \@secondoftwo
 \fi
}%
\providecommand \natexlab [1]{#1}%
\providecommand \enquote  [1]{``#1''}%
\providecommand \bibnamefont  [1]{#1}%
\providecommand \bibfnamefont [1]{#1}%
\providecommand \citenamefont [1]{#1}%
\providecommand \href@noop [0]{\@secondoftwo}%
\providecommand \href [0]{\begingroup \@sanitize@url \@href}%
\providecommand \@href[1]{\@@startlink{#1}\@@href}%
\providecommand \@@href[1]{\endgroup#1\@@endlink}%
\providecommand \@sanitize@url [0]{\catcode `\\12\catcode `\$12\catcode
  `\&12\catcode `\#12\catcode `\^12\catcode `\_12\catcode `\%12\relax}%
\providecommand \@@startlink[1]{}%
\providecommand \@@endlink[0]{}%
\providecommand \url  [0]{\begingroup\@sanitize@url \@url }%
\providecommand \@url [1]{\endgroup\@href {#1}{\urlprefix }}%
\providecommand \urlprefix  [0]{URL }%
\providecommand \Eprint [0]{\href }%
\providecommand \doibase [0]{http://dx.doi.org/}%
\providecommand \selectlanguage [0]{\@gobble}%
\providecommand \bibinfo  [0]{\@secondoftwo}%
\providecommand \bibfield  [0]{\@secondoftwo}%
\providecommand \translation [1]{[#1]}%
\providecommand \BibitemOpen [0]{}%
\providecommand \bibitemStop [0]{}%
\providecommand \bibitemNoStop [0]{.\EOS\space}%
\providecommand \EOS [0]{\spacefactor3000\relax}%
\providecommand \BibitemShut  [1]{\csname bibitem#1\endcsname}%
\let\auto@bib@innerbib\@empty
\bibitem [{\citenamefont {{D'Yakonov}}\ and\ \citenamefont
  {{Perel'}}(1971)}]{Dyakonov}%
  \BibitemOpen
  \bibfield  {author} {\bibinfo {author} {\bibfnamefont {M.~I.}\ \bibnamefont
  {{D'Yakonov}}}\ and\ \bibinfo {author} {\bibfnamefont {V.~I.}\ \bibnamefont
  {{Perel'}}},\ }\href@noop {} {\bibfield  {journal} {\bibinfo  {journal}
  {Soviet Journal of Experimental and Theoretical Physics Letters}\ }\textbf
  {\bibinfo {volume} {13}},\ \bibinfo {pages} {467} (\bibinfo {year}
  {1971})}\BibitemShut {NoStop}%
\bibitem [{\citenamefont {Hirsch}(1999)}]{Hirsch1999}%
  \BibitemOpen
  \bibfield  {author} {\bibinfo {author} {\bibfnamefont {J.~E.}\ \bibnamefont
  {Hirsch}},\ }\href {\doibase 10.1103/PhysRevLett.83.1834} {\bibfield
  {journal} {\bibinfo  {journal} {Phys. Rev. Lett.}\ }\textbf {\bibinfo
  {volume} {83}},\ \bibinfo {pages} {1834} (\bibinfo {year}
  {1999})}\BibitemShut {NoStop}%
\bibitem [{\citenamefont {Kato}\ \emph {et~al.}(2004)\citenamefont {Kato},
  \citenamefont {Myers}, \citenamefont {Gossard},\ and\ \citenamefont
  {Awschalom}}]{reviewSHE}%
  \BibitemOpen
  \bibfield  {author} {\bibinfo {author} {\bibfnamefont {Y.~K.}\ \bibnamefont
  {Kato}}, \bibinfo {author} {\bibfnamefont {R.~C.}\ \bibnamefont {Myers}},
  \bibinfo {author} {\bibfnamefont {A.~C.}\ \bibnamefont {Gossard}}, \ and\
  \bibinfo {author} {\bibfnamefont {D.~D.}\ \bibnamefont {Awschalom}},\ }\href
  {\doibase 10.1126/science.1105514} {\bibfield  {journal} {\bibinfo  {journal}
  {Science}\ }\textbf {\bibinfo {volume} {306}},\ \bibinfo {pages} {1910}
  (\bibinfo {year} {2004})},\ \Eprint
  {http://arxiv.org/abs/http://www.sciencemag.org/content/306/5703/1910.full.pdf}
  {http://www.sciencemag.org/content/306/5703/1910.full.pdf} \BibitemShut
  {NoStop}%
\bibitem [{\citenamefont {Geim}\ and\ \citenamefont
  {Novoselov}(2007)}]{Geim2007}%
  \BibitemOpen
  \bibfield  {author} {\bibinfo {author} {\bibfnamefont {A.~K.}\ \bibnamefont
  {Geim}}\ and\ \bibinfo {author} {\bibfnamefont {K.~S.}\ \bibnamefont
  {Novoselov}},\ }\href {http://dx.doi.org/10.1038/nmat1849} {\bibfield
  {journal} {\bibinfo  {journal} {Nat Mater}\ }\textbf {\bibinfo {volume}
  {6}},\ \bibinfo {pages} {183} (\bibinfo {year} {2007})}\BibitemShut {NoStop}%
\bibitem [{\citenamefont {Kane}\ and\ \citenamefont
  {Mele}(2005{\natexlab{a}})}]{Kane2005}%
  \BibitemOpen
  \bibfield  {author} {\bibinfo {author} {\bibfnamefont {C.~L.}\ \bibnamefont
  {Kane}}\ and\ \bibinfo {author} {\bibfnamefont {E.~J.}\ \bibnamefont
  {Mele}},\ }\href {\doibase 10.1103/PhysRevLett.95.226801} {\bibfield
  {journal} {\bibinfo  {journal} {Phys. Rev. Lett.}\ }\textbf {\bibinfo
  {volume} {95}},\ \bibinfo {pages} {226801} (\bibinfo {year}
  {2005}{\natexlab{a}})}\BibitemShut {NoStop}%
\bibitem [{\citenamefont {König}\ \emph {et~al.}(2007)\citenamefont {König},
  \citenamefont {Wiedmann}, \citenamefont {Brüne}, \citenamefont {Roth},
  \citenamefont {Buhmann}, \citenamefont {Molenkamp}, \citenamefont {Qi},\ and\
  \citenamefont {Zhang}}]{QSHE}%
  \BibitemOpen
  \bibfield  {author} {\bibinfo {author} {\bibfnamefont {M.}~\bibnamefont
  {König}}, \bibinfo {author} {\bibfnamefont {S.}~\bibnamefont {Wiedmann}},
  \bibinfo {author} {\bibfnamefont {C.}~\bibnamefont {Brüne}}, \bibinfo
  {author} {\bibfnamefont {A.}~\bibnamefont {Roth}}, \bibinfo {author}
  {\bibfnamefont {H.}~\bibnamefont {Buhmann}}, \bibinfo {author} {\bibfnamefont
  {L.~W.}\ \bibnamefont {Molenkamp}}, \bibinfo {author} {\bibfnamefont {X.-L.}\
  \bibnamefont {Qi}}, \ and\ \bibinfo {author} {\bibfnamefont {S.-C.}\
  \bibnamefont {Zhang}},\ }\href {\doibase 10.1126/science.1148047} {\bibfield
  {journal} {\bibinfo  {journal} {Science}\ }\textbf {\bibinfo {volume}
  {318}},\ \bibinfo {pages} {766} (\bibinfo {year} {2007})},\ \Eprint
  {http://arxiv.org/abs/http://www.sciencemag.org/content/318/5851/766.full.pdf}
  {http://www.sciencemag.org/content/318/5851/766.full.pdf} \BibitemShut
  {NoStop}%
\bibitem [{\citenamefont {Hasan}\ and\ \citenamefont {Kane}(2010)}]{Kane2010}%
  \BibitemOpen
  \bibfield  {author} {\bibinfo {author} {\bibfnamefont {M.~Z.}\ \bibnamefont
  {Hasan}}\ and\ \bibinfo {author} {\bibfnamefont {C.~L.}\ \bibnamefont
  {Kane}},\ }\href {\doibase 10.1103/RevModPhys.82.3045} {\bibfield  {journal}
  {\bibinfo  {journal} {Rev. Mod. Phys.}\ }\textbf {\bibinfo {volume} {82}},\
  \bibinfo {pages} {3045} (\bibinfo {year} {2010})}\BibitemShut {NoStop}%
\bibitem [{\citenamefont {Kane}\ and\ \citenamefont
  {Mele}(2005{\natexlab{b}})}]{Kane2005b}%
  \BibitemOpen
  \bibfield  {author} {\bibinfo {author} {\bibfnamefont {C.~L.}\ \bibnamefont
  {Kane}}\ and\ \bibinfo {author} {\bibfnamefont {E.~J.}\ \bibnamefont
  {Mele}},\ }\href {\doibase 10.1103/PhysRevLett.95.146802} {\bibfield
  {journal} {\bibinfo  {journal} {Phys. Rev. Lett.}\ }\textbf {\bibinfo
  {volume} {95}},\ \bibinfo {pages} {146802} (\bibinfo {year}
  {2005}{\natexlab{b}})}\BibitemShut {NoStop}%
\bibitem [{\citenamefont {Dyakonov}\ and\ \citenamefont
  {Perel}(1972)}]{Dyakonov2}%
  \BibitemOpen
  \bibfield  {author} {\bibinfo {author} {\bibfnamefont {M.}~\bibnamefont
  {Dyakonov}}\ and\ \bibinfo {author} {\bibfnamefont {V.}~\bibnamefont
  {Perel}},\ }\href@noop {} {\bibfield  {journal} {\bibinfo  {journal} {Sov.
  Phys. Solid State}\ }\textbf {\bibinfo {volume} {13}},\ \bibinfo {pages}
  {3023} (\bibinfo {year} {1972})}\BibitemShut {NoStop}%
\bibitem [{\citenamefont {Gmitra}\ \emph {et~al.}(2013)\citenamefont {Gmitra},
  \citenamefont {Kochan},\ and\ \citenamefont {Fabian}}]{Gmitra2013}%
  \BibitemOpen
  \bibfield  {author} {\bibinfo {author} {\bibfnamefont {M.}~\bibnamefont
  {Gmitra}}, \bibinfo {author} {\bibfnamefont {D.}~\bibnamefont {Kochan}}, \
  and\ \bibinfo {author} {\bibfnamefont {J.}~\bibnamefont {Fabian}},\ }\href
  {\doibase 10.1103/PhysRevLett.110.246602} {\bibfield  {journal} {\bibinfo
  {journal} {Phys. Rev. Lett.}\ }\textbf {\bibinfo {volume} {110}},\ \bibinfo
  {pages} {246602} (\bibinfo {year} {2013})}\BibitemShut {NoStop}%
\bibitem [{\citenamefont {Balakrishnan}\ \emph {et~al.}(2013)\citenamefont
  {Balakrishnan}, \citenamefont {Kok Wai~Koon}, \citenamefont {Jaiswal},
  \citenamefont {Castro~Neto},\ and\ \citenamefont
  {Ozyilmaz}}]{Balakrishnan2013}%
  \BibitemOpen
  \bibfield  {author} {\bibinfo {author} {\bibfnamefont {J.}~\bibnamefont
  {Balakrishnan}}, \bibinfo {author} {\bibfnamefont {G.}~\bibnamefont {Kok
  Wai~Koon}}, \bibinfo {author} {\bibfnamefont {M.}~\bibnamefont {Jaiswal}},
  \bibinfo {author} {\bibfnamefont {A.~H.}\ \bibnamefont {Castro~Neto}}, \ and\
  \bibinfo {author} {\bibfnamefont {B.}~\bibnamefont {Ozyilmaz}},\ }\href
  {http://dx.doi.org/10.1038/nphys2576} {\bibfield  {journal} {\bibinfo
  {journal} {Nat Phys}\ }\textbf {\bibinfo {volume} {9}},\ \bibinfo {pages}
  {284} (\bibinfo {year} {2013})}\BibitemShut {NoStop}%
\bibitem [{\citenamefont {Soltan-Panahi}\ \emph {et~al.}(2011)\citenamefont
  {Soltan-Panahi}, \citenamefont {Struck}, \citenamefont {Hauke}, \citenamefont
  {Bick}, \citenamefont {Plenkers}, \citenamefont {Meineke}, \citenamefont
  {Becker}, \citenamefont {Windpassinger}, \citenamefont {Lewenstein},\ and\
  \citenamefont {Sengstock}}]{hon_atom}%
  \BibitemOpen
  \bibfield  {author} {\bibinfo {author} {\bibfnamefont {P.}~\bibnamefont
  {Soltan-Panahi}}, \bibinfo {author} {\bibfnamefont {J.}~\bibnamefont
  {Struck}}, \bibinfo {author} {\bibfnamefont {P.}~\bibnamefont {Hauke}},
  \bibinfo {author} {\bibfnamefont {A.}~\bibnamefont {Bick}}, \bibinfo {author}
  {\bibfnamefont {W.}~\bibnamefont {Plenkers}}, \bibinfo {author}
  {\bibfnamefont {G.}~\bibnamefont {Meineke}}, \bibinfo {author} {\bibfnamefont
  {C.}~\bibnamefont {Becker}}, \bibinfo {author} {\bibfnamefont
  {P.}~\bibnamefont {Windpassinger}}, \bibinfo {author} {\bibfnamefont
  {M.}~\bibnamefont {Lewenstein}}, \ and\ \bibinfo {author} {\bibfnamefont
  {K.}~\bibnamefont {Sengstock}},\ }\href {http://dx.doi.org/10.1038/nphys1916}
  {\bibfield  {journal} {\bibinfo  {journal} {Nat Phys}\ }\textbf {\bibinfo
  {volume} {7}},\ \bibinfo {pages} {434} (\bibinfo {year} {2011})}\BibitemShut
  {NoStop}%
\bibitem [{\citenamefont {Peleg}\ \emph {et~al.}(2007)\citenamefont {Peleg},
  \citenamefont {Bartal}, \citenamefont {Freedman}, \citenamefont {Manela},
  \citenamefont {Segev},\ and\ \citenamefont {Christodoulides}}]{Peleg2007}%
  \BibitemOpen
  \bibfield  {author} {\bibinfo {author} {\bibfnamefont {O.}~\bibnamefont
  {Peleg}}, \bibinfo {author} {\bibfnamefont {G.}~\bibnamefont {Bartal}},
  \bibinfo {author} {\bibfnamefont {B.}~\bibnamefont {Freedman}}, \bibinfo
  {author} {\bibfnamefont {O.}~\bibnamefont {Manela}}, \bibinfo {author}
  {\bibfnamefont {M.}~\bibnamefont {Segev}}, \ and\ \bibinfo {author}
  {\bibfnamefont {D.~N.}\ \bibnamefont {Christodoulides}},\ }\href {\doibase
  10.1103/PhysRevLett.98.103901} {\bibfield  {journal} {\bibinfo  {journal}
  {Phys. Rev. Lett.}\ }\textbf {\bibinfo {volume} {98}},\ \bibinfo {pages}
  {103901} (\bibinfo {year} {2007})}\BibitemShut {NoStop}%
\bibitem [{\citenamefont {Kuhl}\ \emph {et~al.}(2010)\citenamefont {Kuhl},
  \citenamefont {Barkhofen}, \citenamefont {Tudorovskiy}, \citenamefont
  {St\"ockmann}, \citenamefont {Hossain}, \citenamefont {de~Forges~de Parny},\
  and\ \citenamefont {Mortessagne}}]{Kuhl2010}%
  \BibitemOpen
  \bibfield  {author} {\bibinfo {author} {\bibfnamefont {U.}~\bibnamefont
  {Kuhl}}, \bibinfo {author} {\bibfnamefont {S.}~\bibnamefont {Barkhofen}},
  \bibinfo {author} {\bibfnamefont {T.}~\bibnamefont {Tudorovskiy}}, \bibinfo
  {author} {\bibfnamefont {H.-J.}\ \bibnamefont {St\"ockmann}}, \bibinfo
  {author} {\bibfnamefont {T.}~\bibnamefont {Hossain}}, \bibinfo {author}
  {\bibfnamefont {L.}~\bibnamefont {de~Forges~de Parny}}, \ and\ \bibinfo
  {author} {\bibfnamefont {F.}~\bibnamefont {Mortessagne}},\ }\href {\doibase
  10.1103/PhysRevB.82.094308} {\bibfield  {journal} {\bibinfo  {journal} {Phys.
  Rev. B}\ }\textbf {\bibinfo {volume} {82}},\ \bibinfo {pages} {094308}
  (\bibinfo {year} {2010})}\BibitemShut {NoStop}%
\bibitem [{\citenamefont {Polini}\ \emph {et~al.}(2013)\citenamefont {Polini},
  \citenamefont {Guinea}, \citenamefont {Lewenstein}, \citenamefont
  {Manoharan},\ and\ \citenamefont {Pellegrini}}]{Polini2013}%
  \BibitemOpen
  \bibfield  {author} {\bibinfo {author} {\bibfnamefont {M.}~\bibnamefont
  {Polini}}, \bibinfo {author} {\bibfnamefont {F.}~\bibnamefont {Guinea}},
  \bibinfo {author} {\bibfnamefont {M.}~\bibnamefont {Lewenstein}}, \bibinfo
  {author} {\bibfnamefont {H.~C.}\ \bibnamefont {Manoharan}}, \ and\ \bibinfo
  {author} {\bibfnamefont {V.}~\bibnamefont {Pellegrini}},\ }\href
  {http://dx.doi.org/10.1038/nnano.2013.161} {\bibfield  {journal} {\bibinfo
  {journal} {Nat Nano}\ }\textbf {\bibinfo {volume} {8}},\ \bibinfo {pages}
  {625} (\bibinfo {year} {2013})}\BibitemShut {NoStop}%
\bibitem [{\citenamefont {Kalesaki}\ \emph {et~al.}(2014)\citenamefont
  {Kalesaki}, \citenamefont {Delerue}, \citenamefont {Morais~Smith},
  \citenamefont {Beugeling}, \citenamefont {Allan},\ and\ \citenamefont
  {Vanmaekelbergh}}]{Kalesaki2014}%
  \BibitemOpen
  \bibfield  {author} {\bibinfo {author} {\bibfnamefont {E.}~\bibnamefont
  {Kalesaki}}, \bibinfo {author} {\bibfnamefont {C.}~\bibnamefont {Delerue}},
  \bibinfo {author} {\bibfnamefont {C.}~\bibnamefont {Morais~Smith}}, \bibinfo
  {author} {\bibfnamefont {W.}~\bibnamefont {Beugeling}}, \bibinfo {author}
  {\bibfnamefont {G.}~\bibnamefont {Allan}}, \ and\ \bibinfo {author}
  {\bibfnamefont {D.}~\bibnamefont {Vanmaekelbergh}},\ }\href {\doibase
  10.1103/PhysRevX.4.011010} {\bibfield  {journal} {\bibinfo  {journal} {Phys.
  Rev. X}\ }\textbf {\bibinfo {volume} {4}},\ \bibinfo {pages} {011010}
  (\bibinfo {year} {2014})}\BibitemShut {NoStop}%
\bibitem [{\citenamefont {Jacqmin}\ \emph {et~al.}(2014)\citenamefont
  {Jacqmin}, \citenamefont {Carusotto}, \citenamefont {Sagnes}, \citenamefont
  {Abbarchi}, \citenamefont {Solnyshkov}, \citenamefont {Malpuech},
  \citenamefont {Galopin}, \citenamefont {Lemaitre}, \citenamefont {Bloch},\
  and\ \citenamefont {Amo}}]{Jacqmin2014}%
  \BibitemOpen
  \bibfield  {author} {\bibinfo {author} {\bibfnamefont {T.}~\bibnamefont
  {Jacqmin}}, \bibinfo {author} {\bibfnamefont {I.}~\bibnamefont {Carusotto}},
  \bibinfo {author} {\bibfnamefont {I.}~\bibnamefont {Sagnes}}, \bibinfo
  {author} {\bibfnamefont {M.}~\bibnamefont {Abbarchi}}, \bibinfo {author}
  {\bibfnamefont {D.}~\bibnamefont {Solnyshkov}}, \bibinfo {author}
  {\bibfnamefont {G.}~\bibnamefont {Malpuech}}, \bibinfo {author}
  {\bibfnamefont {E.}~\bibnamefont {Galopin}}, \bibinfo {author} {\bibfnamefont
  {A.}~\bibnamefont {Lemaitre}}, \bibinfo {author} {\bibfnamefont
  {J.}~\bibnamefont {Bloch}}, \ and\ \bibinfo {author} {\bibfnamefont
  {A.}~\bibnamefont {Amo}},\ }\href {\doibase 10.1103/PhysRevLett.112.116402}
  {\bibfield  {journal} {\bibinfo  {journal} {Phys. Rev. Lett.}\ }\textbf
  {\bibinfo {volume} {112}},\ \bibinfo {pages} {116402} (\bibinfo {year}
  {2014})}\BibitemShut {NoStop}%
\bibitem [{\citenamefont {Hafezi}\ \emph {et~al.}(2011)\citenamefont {Hafezi},
  \citenamefont {Demler}, \citenamefont {Lukin},\ and\ \citenamefont
  {Taylor}}]{Hafezi}%
  \BibitemOpen
  \bibfield  {author} {\bibinfo {author} {\bibfnamefont {M.}~\bibnamefont
  {Hafezi}}, \bibinfo {author} {\bibfnamefont {E.}~\bibnamefont {Demler}},
  \bibinfo {author} {\bibfnamefont {M.}~\bibnamefont {Lukin}}, \ and\ \bibinfo
  {author} {\bibfnamefont {J.}~\bibnamefont {Taylor}},\ }\href@noop {}
  {\bibfield  {journal} {\bibinfo  {journal} {Nature Physics}\ }\textbf
  {\bibinfo {volume} {7}},\ \bibinfo {pages} {907} (\bibinfo {year}
  {2011})}\BibitemShut {NoStop}%
\bibitem [{\citenamefont {Umucalilar}\ and\ \citenamefont
  {Carusotto}(2012)}]{Umucalilar}%
  \BibitemOpen
  \bibfield  {author} {\bibinfo {author} {\bibfnamefont {R.}~\bibnamefont
  {Umucalilar}}\ and\ \bibinfo {author} {\bibfnamefont {I.}~\bibnamefont
  {Carusotto}},\ }\href@noop {} {\bibfield  {journal} {\bibinfo  {journal}
  {Phys. Rev. Lett.}\ }\textbf {\bibinfo {volume} {108}},\ \bibinfo {pages}
  {206809} (\bibinfo {year} {2012})}\BibitemShut {NoStop}%
\bibitem [{\citenamefont {Lin}\ \emph {et~al.}(2009)\citenamefont {Lin},
  \citenamefont {Compton}, \citenamefont {Jimenez-Garcia}, \citenamefont
  {Porto},\ and\ \citenamefont {Spielman}}]{Lin1}%
  \BibitemOpen
  \bibfield  {author} {\bibinfo {author} {\bibfnamefont {Y.-J.}\ \bibnamefont
  {Lin}}, \bibinfo {author} {\bibfnamefont {R.~L.}\ \bibnamefont {Compton}},
  \bibinfo {author} {\bibfnamefont {K.}~\bibnamefont {Jimenez-Garcia}},
  \bibinfo {author} {\bibfnamefont {J.~V.}\ \bibnamefont {Porto}}, \ and\
  \bibinfo {author} {\bibfnamefont {I.~B.}\ \bibnamefont {Spielman}},\ }\href
  {http://dx.doi.org/10.1038/nature08609} {\bibfield  {journal} {\bibinfo
  {journal} {Nature}\ }\textbf {\bibinfo {volume} {462}},\ \bibinfo {pages}
  {628} (\bibinfo {year} {2009})}\BibitemShut {NoStop}%
\bibitem [{\citenamefont {Lin}\ \emph {et~al.}(2011)\citenamefont {Lin},
  \citenamefont {Jimenez-Garcia},\ and\ \citenamefont {Spielman}}]{Lin2}%
  \BibitemOpen
  \bibfield  {author} {\bibinfo {author} {\bibfnamefont {Y.-J.}\ \bibnamefont
  {Lin}}, \bibinfo {author} {\bibfnamefont {K.}~\bibnamefont {Jimenez-Garcia}},
  \ and\ \bibinfo {author} {\bibfnamefont {I.~B.}\ \bibnamefont {Spielman}},\
  }\href {http://dx.doi.org/10.1038/nature09887} {\bibfield  {journal}
  {\bibinfo  {journal} {Nature}\ }\textbf {\bibinfo {volume} {471}},\ \bibinfo
  {pages} {83} (\bibinfo {year} {2011})}\BibitemShut {NoStop}%
\bibitem [{\citenamefont {Rechtsman}\ \emph {et~al.}(2013)\citenamefont
  {Rechtsman}, \citenamefont {Zeuner}, \citenamefont {Plotnik}, \citenamefont
  {Lumer}, \citenamefont {Podolsky}, \citenamefont {Dreisow}, \citenamefont
  {Nolte}, \citenamefont {Segev},\ and\ \citenamefont
  {Szameit}}]{Rechtsman2013}%
  \BibitemOpen
  \bibfield  {author} {\bibinfo {author} {\bibfnamefont {M.}~\bibnamefont
  {Rechtsman}}, \bibinfo {author} {\bibfnamefont {J.}~\bibnamefont {Zeuner}},
  \bibinfo {author} {\bibfnamefont {Y.}~\bibnamefont {Plotnik}}, \bibinfo
  {author} {\bibfnamefont {Y.}~\bibnamefont {Lumer}}, \bibinfo {author}
  {\bibfnamefont {D.}~\bibnamefont {Podolsky}}, \bibinfo {author}
  {\bibfnamefont {F.}~\bibnamefont {Dreisow}}, \bibinfo {author} {\bibfnamefont
  {S.}~\bibnamefont {Nolte}}, \bibinfo {author} {\bibfnamefont
  {M.}~\bibnamefont {Segev}}, \ and\ \bibinfo {author} {\bibfnamefont
  {A.}~\bibnamefont {Szameit}},\ }\href@noop {} {\bibfield  {journal} {\bibinfo
   {journal} {Nature}\ }\textbf {\bibinfo {volume} {496}},\ \bibinfo {pages}
  {196} (\bibinfo {year} {2013})}\BibitemShut {NoStop}%
\bibitem [{\citenamefont {Won}(2011)}]{Won2011}%
  \BibitemOpen
  \bibfield  {author} {\bibinfo {author} {\bibfnamefont {R.}~\bibnamefont
  {Won}},\ }\href@noop {} {\bibfield  {journal} {\bibinfo  {journal} {Nature
  Photonics}\ }\textbf {\bibinfo {volume} {5}},\ \bibinfo {pages} {512}
  (\bibinfo {year} {2011})}\BibitemShut {NoStop}%
\bibitem [{\citenamefont {Chong}(2013)}]{Chong2013}%
  \BibitemOpen
  \bibfield  {author} {\bibinfo {author} {\bibfnamefont {Y.}~\bibnamefont
  {Chong}},\ }\href {http://dx.doi.org/10.1038/496173a} {\bibfield  {journal}
  {\bibinfo  {journal} {Nature}\ }\textbf {\bibinfo {volume} {496}},\ \bibinfo
  {pages} {173} (\bibinfo {year} {2013})}\BibitemShut {NoStop}%
\bibitem [{\citenamefont {Khanikaev}\ \emph {et~al.}(2013)\citenamefont
  {Khanikaev}, \citenamefont {Mousavi}, \citenamefont {Tse}, \citenamefont
  {Kargarian}, \citenamefont {MacDonald},\ and\ \citenamefont
  {Shvets}}]{Khanikaev}%
  \BibitemOpen
  \bibfield  {author} {\bibinfo {author} {\bibfnamefont {A.}~\bibnamefont
  {Khanikaev}}, \bibinfo {author} {\bibfnamefont {S.}~\bibnamefont {Mousavi}},
  \bibinfo {author} {\bibfnamefont {W.-K.}\ \bibnamefont {Tse}}, \bibinfo
  {author} {\bibfnamefont {M.}~\bibnamefont {Kargarian}}, \bibinfo {author}
  {\bibfnamefont {A.}~\bibnamefont {MacDonald}}, \ and\ \bibinfo {author}
  {\bibfnamefont {G.}~\bibnamefont {Shvets}},\ }\href@noop {} {\bibfield
  {journal} {\bibinfo  {journal} {Nature Materials}\ }\textbf {\bibinfo
  {volume} {12}},\ \bibinfo {pages} {233} (\bibinfo {year} {2013})}\BibitemShut
  {NoStop}%
\bibitem [{\citenamefont {Shelykh}\ \emph {et~al.}(2010)\citenamefont
  {Shelykh}, \citenamefont {Kavokin}, \citenamefont {Rubo}, \citenamefont
  {Liew},\ and\ \citenamefont {Malpuech}}]{Shelykh2010}%
  \BibitemOpen
  \bibfield  {author} {\bibinfo {author} {\bibfnamefont {I.~A.}\ \bibnamefont
  {Shelykh}}, \bibinfo {author} {\bibfnamefont {A.~V.}\ \bibnamefont
  {Kavokin}}, \bibinfo {author} {\bibfnamefont {Y.~G.}\ \bibnamefont {Rubo}},
  \bibinfo {author} {\bibfnamefont {T.~C.~H.}\ \bibnamefont {Liew}}, \ and\
  \bibinfo {author} {\bibfnamefont {G.}~\bibnamefont {Malpuech}},\ }\href
  {http://stacks.iop.org/0268-1242/25/i=1/a=013001} {\bibfield  {journal}
  {\bibinfo  {journal} {Semiconductor Science and Technology}\ }\textbf
  {\bibinfo {volume} {25}},\ \bibinfo {pages} {013001} (\bibinfo {year}
  {2010})}\BibitemShut {NoStop}%
\bibitem [{\citenamefont {Kavokin}\ \emph {et~al.}(2005)\citenamefont
  {Kavokin}, \citenamefont {Malpuech},\ and\ \citenamefont
  {Glazov}}]{Kavokin2005}%
  \BibitemOpen
  \bibfield  {author} {\bibinfo {author} {\bibfnamefont {A.}~\bibnamefont
  {Kavokin}}, \bibinfo {author} {\bibfnamefont {G.}~\bibnamefont {Malpuech}}, \
  and\ \bibinfo {author} {\bibfnamefont {M.}~\bibnamefont {Glazov}},\ }\href
  {\doibase 10.1103/PhysRevLett.95.136601} {\bibfield  {journal} {\bibinfo
  {journal} {Phys. Rev. Lett.}\ }\textbf {\bibinfo {volume} {95}},\ \bibinfo
  {pages} {136601} (\bibinfo {year} {2005})}\BibitemShut {NoStop}%
\bibitem [{\citenamefont {Leyder}\ \emph {et~al.}(2007)\citenamefont {Leyder},
  \citenamefont {Romanelli}, \citenamefont {Karr}, \citenamefont {Giacobino},
  \citenamefont {Liew}, \citenamefont {Glazov}, \citenamefont {Kavokin},
  \citenamefont {Malpuech},\ and\ \citenamefont {Bramati}}]{Leyder2007}%
  \BibitemOpen
  \bibfield  {author} {\bibinfo {author} {\bibfnamefont {C.}~\bibnamefont
  {Leyder}}, \bibinfo {author} {\bibfnamefont {M.}~\bibnamefont {Romanelli}},
  \bibinfo {author} {\bibfnamefont {J.~P.}\ \bibnamefont {Karr}}, \bibinfo
  {author} {\bibfnamefont {E.}~\bibnamefont {Giacobino}}, \bibinfo {author}
  {\bibfnamefont {T.~C.~H.}\ \bibnamefont {Liew}}, \bibinfo {author}
  {\bibfnamefont {M.~M.}\ \bibnamefont {Glazov}}, \bibinfo {author}
  {\bibfnamefont {A.~V.}\ \bibnamefont {Kavokin}}, \bibinfo {author}
  {\bibfnamefont {G.}~\bibnamefont {Malpuech}}, \ and\ \bibinfo {author}
  {\bibfnamefont {A.}~\bibnamefont {Bramati}},\ }\href
  {http://dx.doi.org/10.1038/nphys676} {\bibfield  {journal} {\bibinfo
  {journal} {Nat Phys}\ }\textbf {\bibinfo {volume} {3}},\ \bibinfo {pages}
  {628} (\bibinfo {year} {2007})}\BibitemShut {NoStop}%
\bibitem [{\citenamefont {Hivet}\ \emph {et~al.}(2012)\citenamefont {Hivet},
  \citenamefont {Flayac}, \citenamefont {Solnyshkov}, \citenamefont {Tanese},
  \citenamefont {Boulier}, \citenamefont {Andreoli}, \citenamefont {Giacobino},
  \citenamefont {Bloch}, \citenamefont {Bramati}, \citenamefont {Malpuech},\
  and\ \citenamefont {Amo}}]{Hivet}%
  \BibitemOpen
  \bibfield  {author} {\bibinfo {author} {\bibfnamefont {R.}~\bibnamefont
  {Hivet}}, \bibinfo {author} {\bibfnamefont {H.}~\bibnamefont {Flayac}},
  \bibinfo {author} {\bibfnamefont {D.~D.}\ \bibnamefont {Solnyshkov}},
  \bibinfo {author} {\bibfnamefont {D.}~\bibnamefont {Tanese}}, \bibinfo
  {author} {\bibfnamefont {T.}~\bibnamefont {Boulier}}, \bibinfo {author}
  {\bibfnamefont {D.}~\bibnamefont {Andreoli}}, \bibinfo {author}
  {\bibfnamefont {E.}~\bibnamefont {Giacobino}}, \bibinfo {author}
  {\bibfnamefont {J.}~\bibnamefont {Bloch}}, \bibinfo {author} {\bibfnamefont
  {A.}~\bibnamefont {Bramati}}, \bibinfo {author} {\bibfnamefont
  {G.}~\bibnamefont {Malpuech}}, \ and\ \bibinfo {author} {\bibfnamefont
  {A.}~\bibnamefont {Amo}},\ }\href {http://dx.doi.org/10.1038/nphys2406}
  {\bibfield  {journal} {\bibinfo  {journal} {Nat Phys}\ }\textbf {\bibinfo
  {volume} {8}},\ \bibinfo {pages} {724} (\bibinfo {year} {2012})}\BibitemShut
  {NoStop}%
\bibitem [{\citenamefont {Bramwell}(2012)}]{Bramwell2012}%
  \BibitemOpen
  \bibfield  {author} {\bibinfo {author} {\bibfnamefont {S.~T.}\ \bibnamefont
  {Bramwell}},\ }\href {http://dx.doi.org/10.1038/nphys2412} {\bibfield
  {journal} {\bibinfo  {journal} {Nat Phys}\ }\textbf {\bibinfo {volume} {8}},\
  \bibinfo {pages} {703} (\bibinfo {year} {2012})}\BibitemShut {NoStop}%
\bibitem [{\citenamefont {Solnyshkov}\ \emph {et~al.}(2012)\citenamefont
  {Solnyshkov}, \citenamefont {Flayac},\ and\ \citenamefont
  {Malpuech}}]{Solnyshkov2013}%
  \BibitemOpen
  \bibfield  {author} {\bibinfo {author} {\bibfnamefont {D.~D.}\ \bibnamefont
  {Solnyshkov}}, \bibinfo {author} {\bibfnamefont {H.}~\bibnamefont {Flayac}},
  \ and\ \bibinfo {author} {\bibfnamefont {G.}~\bibnamefont {Malpuech}},\
  }\href {\doibase 10.1103/PhysRevB.85.073105} {\bibfield  {journal} {\bibinfo
  {journal} {Phys. Rev. B}\ }\textbf {\bibinfo {volume} {85}},\ \bibinfo
  {pages} {073105} (\bibinfo {year} {2012})}\BibitemShut {NoStop}%
\bibitem [{\citenamefont {Tercas}\ \emph {et~al.}(2014)\citenamefont {Tercas},
  \citenamefont {Flayac}, \citenamefont {Solnyshkov},\ and\ \citenamefont
  {Malpuech}}]{Tercas2014}%
  \BibitemOpen
  \bibfield  {author} {\bibinfo {author} {\bibfnamefont {H.}~\bibnamefont
  {Tercas}}, \bibinfo {author} {\bibfnamefont {H.}~\bibnamefont {Flayac}},
  \bibinfo {author} {\bibfnamefont {D.}~\bibnamefont {Solnyshkov},
  \bibfnamefont {D.}}, \ and\ \bibinfo {author} {\bibfnamefont
  {G.}~\bibnamefont {Malpuech}},\ }\href {\doibase
  10.1103/PhysRevLett.112.066402} {\bibfield  {journal} {\bibinfo  {journal}
  {Phys. Rev. Lett.}\ }\textbf {\bibinfo {volume} {112}},\ \bibinfo {pages}
  {066402} (\bibinfo {year} {2014})}\BibitemShut {NoStop}%
\bibitem [{\citenamefont {McCann}\ and\ \citenamefont
  {Fal'ko}(2006)}]{McCann2006}%
  \BibitemOpen
  \bibfield  {author} {\bibinfo {author} {\bibfnamefont {E.}~\bibnamefont
  {McCann}}\ and\ \bibinfo {author} {\bibfnamefont {V.~I.}\ \bibnamefont
  {Fal'ko}},\ }\href {\doibase 10.1103/PhysRevLett.96.086805} {\bibfield
  {journal} {\bibinfo  {journal} {Phys. Rev. Lett.}\ }\textbf {\bibinfo
  {volume} {96}},\ \bibinfo {pages} {086805} (\bibinfo {year}
  {2006})}\BibitemShut {NoStop}%
\bibitem [{\citenamefont {Rakyta}\ \emph {et~al.}(2010)\citenamefont {Rakyta},
  \citenamefont {Korm\'anyos},\ and\ \citenamefont {Cserti}}]{Rakyta2010}%
  \BibitemOpen
  \bibfield  {author} {\bibinfo {author} {\bibfnamefont {P.}~\bibnamefont
  {Rakyta}}, \bibinfo {author} {\bibfnamefont {A.}~\bibnamefont {Korm\'anyos}},
  \ and\ \bibinfo {author} {\bibfnamefont {J.}~\bibnamefont {Cserti}},\ }\href
  {\doibase 10.1103/PhysRevB.82.113405} {\bibfield  {journal} {\bibinfo
  {journal} {Phys. Rev. B}\ }\textbf {\bibinfo {volume} {82}},\ \bibinfo
  {pages} {113405} (\bibinfo {year} {2010})}\BibitemShut {NoStop}%
\bibitem [{\citenamefont {Dresselhaus}(1974)}]{Dresselhaus1974}%
  \BibitemOpen
  \bibfield  {author} {\bibinfo {author} {\bibfnamefont {G.}~\bibnamefont
  {Dresselhaus}},\ }\href {\doibase 10.1103/PhysRevB.10.3602} {\bibfield
  {journal} {\bibinfo  {journal} {Phys. Rev. B}\ }\textbf {\bibinfo {volume}
  {10}},\ \bibinfo {pages} {3602} (\bibinfo {year} {1974})}\BibitemShut
  {NoStop}%
\bibitem [{\citenamefont {Lifshitz}(1960)}]{Lifshitz1960}%
  \BibitemOpen
  \bibfield  {author} {\bibinfo {author} {\bibfnamefont {L.}~\bibnamefont
  {Lifshitz}},\ }\href@noop {} {\bibfield  {journal} {\bibinfo  {journal} {Zh.
  Eksp. Teor. Fiz.}\ }\textbf {\bibinfo {volume} {38}},\ \bibinfo {pages}
  {1565} (\bibinfo {year} {1960})}\BibitemShut {NoStop}%
\bibitem [{sup()}]{suppl}%
  \BibitemOpen
  \href@noop {} {}\bibinfo {note} {See Supplemental Material at [URL will be
  inserted by publisher] for the discussion how tunneling is affected by the
  TE-TM splitting.}\BibitemShut {Stop}%
\bibitem [{\citenamefont {Sala}\ \emph {et~al.}(2014)\citenamefont {Sala},
  \citenamefont {Solnyshkov}, \citenamefont {Carusotto}, \citenamefont
  {Jacqmin}, \citenamefont {Lemaitre}, \citenamefont {Tercas}, \citenamefont
  {Nalitov}, \citenamefont {Galopin}, \citenamefont {Sagnes}, \citenamefont
  {Bloch}, \citenamefont {Malpuech},\ and\ \citenamefont {Amo}}]{Vera}%
  \BibitemOpen
  \bibfield  {author} {\bibinfo {author} {\bibfnamefont {V.}~\bibnamefont
  {Sala}}, \bibinfo {author} {\bibfnamefont {D.}~\bibnamefont {Solnyshkov}},
  \bibinfo {author} {\bibfnamefont {I.}~\bibnamefont {Carusotto}}, \bibinfo
  {author} {\bibfnamefont {T.}~\bibnamefont {Jacqmin}}, \bibinfo {author}
  {\bibfnamefont {A.}~\bibnamefont {Lemaitre}}, \bibinfo {author}
  {\bibfnamefont {H.}~\bibnamefont {Tercas}}, \bibinfo {author} {\bibfnamefont
  {A.}~\bibnamefont {Nalitov}}, \bibinfo {author} {\bibfnamefont
  {E.}~\bibnamefont {Galopin}}, \bibinfo {author} {\bibfnamefont
  {I.}~\bibnamefont {Sagnes}}, \bibinfo {author} {\bibfnamefont
  {J.}~\bibnamefont {Bloch}}, \bibinfo {author} {\bibfnamefont
  {G.}~\bibnamefont {Malpuech}}, \ and\ \bibinfo {author} {\bibfnamefont
  {A.}~\bibnamefont {Amo}},\ }\href@noop {} {\bibfield  {journal} {\bibinfo
  {journal} {submitted to Nature Physics}\ } (\bibinfo {year}
  {2014})}\BibitemShut {NoStop}%
\bibitem [{\citenamefont {Wallace}(1947)}]{Wallace}%
  \BibitemOpen
  \bibfield  {author} {\bibinfo {author} {\bibfnamefont {P.~R.}\ \bibnamefont
  {Wallace}},\ }\href {\doibase 10.1103/PhysRev.71.622} {\bibfield  {journal}
  {\bibinfo  {journal} {Phys. Rev.}\ }\textbf {\bibinfo {volume} {71}},\
  \bibinfo {pages} {622} (\bibinfo {year} {1947})}\BibitemShut {NoStop}%
\bibitem [{\citenamefont {Liu}\ \emph {et~al.}(2012)\citenamefont {Liu},
  \citenamefont {Bundesmann},\ and\ \citenamefont {Richter}}]{Liu2012}%
  \BibitemOpen
  \bibfield  {author} {\bibinfo {author} {\bibfnamefont {M.-H.}\ \bibnamefont
  {Liu}}, \bibinfo {author} {\bibfnamefont {J.}~\bibnamefont {Bundesmann}}, \
  and\ \bibinfo {author} {\bibfnamefont {K.}~\bibnamefont {Richter}},\
  }\href@noop {} {\bibfield  {journal} {\bibinfo  {journal} {Phys. Rev. B}\
  }\textbf {\bibinfo {volume} {85}},\ \bibinfo {pages} {085406} (\bibinfo
  {year} {2012})}\BibitemShut {NoStop}%
\bibitem [{\citenamefont {Flayac}\ \emph {et~al.}(2013)\citenamefont {Flayac},
  \citenamefont {Solnyshkov}, \citenamefont {Shelykh},\ and\ \citenamefont
  {Malpuech}}]{NOSHE}%
  \BibitemOpen
  \bibfield  {author} {\bibinfo {author} {\bibfnamefont {H.}~\bibnamefont
  {Flayac}}, \bibinfo {author} {\bibfnamefont {D.~D.}\ \bibnamefont
  {Solnyshkov}}, \bibinfo {author} {\bibfnamefont {I.~A.}\ \bibnamefont
  {Shelykh}}, \ and\ \bibinfo {author} {\bibfnamefont {G.}~\bibnamefont
  {Malpuech}},\ }\href {\doibase 10.1103/PhysRevLett.110.016404} {\bibfield
  {journal} {\bibinfo  {journal} {Phys. Rev. Lett.}\ }\textbf {\bibinfo
  {volume} {110}},\ \bibinfo {pages} {016404} (\bibinfo {year}
  {2013})}\BibitemShut {NoStop}%
\bibitem [{\citenamefont {Vishnevsky}\ \emph {et~al.}(2013)\citenamefont
  {Vishnevsky}, \citenamefont {Flayac}, \citenamefont {Nalitov}, \citenamefont
  {Solnyshkov}, \citenamefont {Gippius},\ and\ \citenamefont
  {Malpuech}}]{Vishnevsky2013}%
  \BibitemOpen
  \bibfield  {author} {\bibinfo {author} {\bibfnamefont {D.~V.}\ \bibnamefont
  {Vishnevsky}}, \bibinfo {author} {\bibfnamefont {H.}~\bibnamefont {Flayac}},
  \bibinfo {author} {\bibfnamefont {A.~V.}\ \bibnamefont {Nalitov}}, \bibinfo
  {author} {\bibfnamefont {D.~D.}\ \bibnamefont {Solnyshkov}}, \bibinfo
  {author} {\bibfnamefont {N.~A.}\ \bibnamefont {Gippius}}, \ and\ \bibinfo
  {author} {\bibfnamefont {G.}~\bibnamefont {Malpuech}},\ }\href {\doibase
  10.1103/PhysRevLett.110.246404} {\bibfield  {journal} {\bibinfo  {journal}
  {Phys. Rev. Lett.}\ }\textbf {\bibinfo {volume} {110}},\ \bibinfo {pages}
  {246404} (\bibinfo {year} {2013})}\BibitemShut {NoStop}%
\bibitem [{\citenamefont {Haldane}\ and\ \citenamefont
  {Raghu}(2008)}]{Haldane2008}%
  \BibitemOpen
  \bibfield  {author} {\bibinfo {author} {\bibfnamefont {F.~D.~M.}\
  \bibnamefont {Haldane}}\ and\ \bibinfo {author} {\bibfnamefont
  {S.}~\bibnamefont {Raghu}},\ }\href {\doibase 10.1103/PhysRevLett.100.013904}
  {\bibfield  {journal} {\bibinfo  {journal} {Phys. Rev. Lett.}\ }\textbf
  {\bibinfo {volume} {100}},\ \bibinfo {pages} {013904} (\bibinfo {year}
  {2008})}\BibitemShut {NoStop}%
\bibitem [{\citenamefont {Wang}\ \emph {et~al.}(2009)\citenamefont {Wang},
  \citenamefont {Chong}, \citenamefont {Joannopoulos},\ and\ \citenamefont
  {Soljacic}}]{Soljacic09}%
  \BibitemOpen
  \bibfield  {author} {\bibinfo {author} {\bibfnamefont {Z.}~\bibnamefont
  {Wang}}, \bibinfo {author} {\bibfnamefont {Y.}~\bibnamefont {Chong}},
  \bibinfo {author} {\bibfnamefont {J.~D.}\ \bibnamefont {Joannopoulos}}, \
  and\ \bibinfo {author} {\bibfnamefont {M.}~\bibnamefont {Soljacic}},\ }\href
  {http://dx.doi.org/10.1038/nature08293} {\bibfield  {journal} {\bibinfo
  {journal} {Nature}\ }\textbf {\bibinfo {volume} {461}},\ \bibinfo {pages}
  {772} (\bibinfo {year} {2009})}\BibitemShut {NoStop}%
\bibitem [{\citenamefont {Lumer}\ \emph {et~al.}(2013)\citenamefont {Lumer},
  \citenamefont {Plotnik}, \citenamefont {Rechtsman},\ and\ \citenamefont
  {Segev}}]{Segev}%
  \BibitemOpen
  \bibfield  {author} {\bibinfo {author} {\bibfnamefont {Y.}~\bibnamefont
  {Lumer}}, \bibinfo {author} {\bibfnamefont {Y.}~\bibnamefont {Plotnik}},
  \bibinfo {author} {\bibfnamefont {M.~C.}\ \bibnamefont {Rechtsman}}, \ and\
  \bibinfo {author} {\bibfnamefont {M.}~\bibnamefont {Segev}},\ }\href
  {\doibase 10.1103/PhysRevLett.111.243905} {\bibfield  {journal} {\bibinfo
  {journal} {Phys. Rev. Lett.}\ }\textbf {\bibinfo {volume} {111}},\ \bibinfo
  {pages} {243905} (\bibinfo {year} {2013})}\BibitemShut {NoStop}%
\bibitem [{\citenamefont {Carusotto}\ and\ \citenamefont
  {Ciuti}(2013)}]{Carusotto2013}%
  \BibitemOpen
  \bibfield  {author} {\bibinfo {author} {\bibfnamefont {I.}~\bibnamefont
  {Carusotto}}\ and\ \bibinfo {author} {\bibfnamefont {C.}~\bibnamefont
  {Ciuti}},\ }\href {\doibase 10.1103/RevModPhys.85.299} {\bibfield  {journal}
  {\bibinfo  {journal} {Rev. Mod. Phys.}\ }\textbf {\bibinfo {volume} {85}},\
  \bibinfo {pages} {299} (\bibinfo {year} {2013})}\BibitemShut {NoStop}%
\end{thebibliography}%

\end{document}